# Framing the Hacker: Media Representations and Public Discourse in Germany

Raphael Morisco
*Independent scholar*
*(Dr. phil., Universität Bielefeld; formerly at Karlsruhe Institute of Technology)*

**Abstract:**

This paper examines how the figure of the hacker is portrayed in German mainstream media and explores the impact of media framing on public discourse. Through a longitudinal content analysis of 301 articles from four of the most widely circulated German newspapers (Die Zeit, Süddeutsche Zeitung, Bild, and Der Spiegel), the study covers reporting between January 2017 and January 2020. The results reveal a strong predominance of negative connotations and dramatizing frames that link hackers to criminality, national security threats, and digital warfare. Drawing on media effects theory, scandalization mechanisms, and constructivist media theory, the article shows how media representations co-construct public perceptions of IT-related risks. The analysis emphasizes the role of agenda setting, framing, and media reality in shaping societal narratives around hackers. The study concludes by reflecting on the broader implications for IT security education and the sociopolitical challenges posed by distorted representations of digital actors.

# Introduction

> "Knowledge of technology […] becomes the dividing line between the hacker and the typical computer user" [1, p. 54].

The examination of hackers, regarded as social figures, allows for the depiction of societal perspectives on IT security issues and the public perception of perceived threats.[1] This is because social figures "are temporally bound historical characters, through which a specific view of contemporary society can be projected" [4, p. 8]. Furthermore, hackers "occupy a pivotal position in cultural debates, media, and social discourses of Western modernity" [4, p. 9]. According to Stephan Moebius and Markus Schroer, hackers are among the "most frequently occurring social figures in cultural and political discussions" [4, p. 9].

The following analysis examines various perspectives to both elucidate the transmedia portrayal of hackers and approximate public perception. Chapter 2 (Media Representation: Initial Perspectives and Fundamentals) begins by exploring the depiction of hackers in popular culture and the shifting manifestations of technological fear, thereby providing an initial media portrayal. Additionally, concepts such as agenda setting and framing are contextualized, along with preliminary discussions on the mechanisms of scandalization in journalism and aspects of media reality. Subsection 2.1. (Mechanisms of Scandalization) outlines the relevant characteristics and tools essential for analyzing the portrayal of hackers in media reality. Following this, and prompted by the frequent use of the term 'media reality' itself, subsection 2.2 (Media Reality) briefly delves into the theoretical interpretation and associated challenges to better contextualize the term.

In Chapter 3 (Media Representation: Data Analysis), a longitudinal study is conducted to examine the media portrayal of hackers in more detail. This involves an explanation of the dataset itself (outlined in Subsection 3.1 (Explanation of Data Derivation for Longitudinal Analysis)) and a quantitative content analysis of the data collection (Section 3.2 (Content Analysis)). Subsequently, the focus shifts to exploring the formation and structure of the media reality image, which is investigated in Subsection 3.3 (Media Frames as a Snapshot of Media Reality from 2017 to 2020). From a content analytical and social scientific perspective, potential implications of the longitudinal study are suggested, briefly discussed in Subsection 3.4 (New Potential Risks and Societal Challenges). Finally, a transition to educational research is made by examining three studies in Chapter 4 (Youth and IT Security), which offer a condensed perspective on how young individuals engage with the topic of IT security through a secondary analysis.

## 2. Media Representation: Initial Perspectives and Fundamentals

According to Douglas Thomas, hackers first captured public attention in the 1980s. This was predominantly exemplified by David Lightman, the protagonist of the hacker thriller *WarGames* (1983) [1, p. xxii]. Thomas suggests that this film served as inspiration for an entire generation of youth to pursue hacking. Another influential figure emerged in the late 1980s in the form of Robert Morris, who unleashed one of the first internet worms (Morris worm),[2] causing widespread disruption across the network [1, p. xxii]. According to Thomas, these two figures had a significant impact on shaping the hacker culture and its portrayal in the media [1, p. xxii]. In his analysis of the hacker culture, he repeatedly criticizes the depictions of hackers in journalistic media, legal proceedings, and popular culture, noting that they tell us more about contemporary cultural attitudes and fears towards new technology than about the hacker culture or the act of hacking itself [1, xx, 9]. Furthermore, Thomas asserts that while these media portrayals of hackers can offer insight into contemporary concerns about new technology, they serve to conceal a sophisticated subculture formed by the hackers themselves [1, p. xx]. He views the image of hackers as a blend of high-tech wizardry and criminality, as hackers are often depicted as criminals both in the media and in the public imagination [1, 5ff]. In this context, hackers are predominantly portrayed in the media as the source of many evils in the highly technological realm of computers. This ranges from computer espionage and digital break-ins to the creation of computer viruses. Hackers are repeatedly presented as the dangerous counterpart to the computer revolution [1, 5ff].

---

[1] A concise and nuanced overview of the term *hacker* can be found in: [2]. For an insider perspective on hacking and hackers, focusing on a small group of individuals who self-identify as hackers, refer to: [3].
[2] On the history of the Morris computer worm, refer to [1, pp. 27-29].

In the portrayal of hacking in popular culture through films, it can be observed that in movies where hackers play a central role, they are almost always depicted as outlaws or criminals. However, the significance of criminality is negotiated within the narratives themselves. Thomas cites the films WarGames (1983), Sneakers (1992), Hackers (1995), and The Net (1995) as examples [1, p. 51]. In these films, hackers are often positioned as petty criminals in relation to a larger scale of crime and injustice perpetuated either by the government, the military, or corporate interests [1, p. 51]. Building on this, Thomas notes that hackers in films where they serve as central protagonists tend to act similarly to the real ethical code of hackers,[3] never pursuing significant personal financial gain but rather preferring to engage in exploration, pranks, personal amusement, or devising ways to improve their local conditions [1, p. 51]. Unlike technology itself, which is almost exclusively connoted as malevolent, Thomas portrays the hacker as an indeterminate character, simultaneously a hero and anti-hero, both the cause and the remedy for social crises [1, p. 52]. From a narrative theory perspective, it becomes evident that these narratives are dichotomous. Thomas argues that, on the one hand, each narrative brings about a certain shift, necessitating hackers to step in for topics of significant cultural concern, exemplified here by the loss of body and identity, the threat of violence, and the fear of national security threats [1, p. 216]. On the other hand, these narratives of fear provide justification for the increased use of technology aimed at creating both a panoptic virtual space and depriving individuals within this virtual space of the secrecy that binds their virtual identity to their physical body [1, p. 216]. In his exploration of the transmedia and cultural portrayal of hackers, Thomas also delves into the realm of jurisprudence, analyzing the significant implications of journalistic characterization as criminal hackers using the examples of Kevin Mitnick and Kevin Poulsen [1, pp. 175-219]. As evidenced by the cases of these hackers, considerable effort is devoted to crafting narratives and images of hackers, aimed at facilitating broader social shifts in fears [1, pp. 175-219]. As part of this representation system, law enforcement, journalistic and artistic media, as well as the state, play a significant role in shaping the manner and style of hacker portrayal [1, p. 219]. Thomas's elaboration precisely illustrates, through the examples of the aforementioned hackers and their activities, how fear of more general social issues is transmedially shifted onto the figure of the hacker.[4]

The challenge of shaping media reality that arises from this shift was pointed out by Espey and Rudinger in the late 1990s regarding "IT security from a psychological perspective" in journalism. They critique the dearth of expert journalists capable of reporting on IT topics with depth and expertise across print and visual media platforms [7, p. 179]. Expanding on this, they argue that "the prevailing sensationalism in journalism employs an entirely incomprehensible agenda setting, resulting in an unrealistic portrayal of the objective risks associated with IT" [7, p. 179]. It is notable that in media and communication studies, *agenda setting* is a term within the lexicon of media effects research, along with *framing*. According to Heinz Bonfadelli, agenda setting and framing differ in focus: "agenda-setting theory concentrates on the frequency and intensity of media coverage of topics or issues and their impacts" [8, p. 188], while "the framing perspective also explores how these topics are reported in the media from various perspectives" [8, p. 188]. In essence, and stated neutrally: media outlets vary in their selection and prioritization of topics (agenda setting), and they also offer diverse perspectives and interpretations of these topics (framing) [8, pp. 173-200]. Espey and Rudinger also point out the discrepancy between factual knowledge and ignorance, implicitly engaging with the Agenda-Setting Theory in their study. Among their critiques is the tendency for complex information to be hastily omitted from news prioritization, resulting in the dominance of "alarmist and sensational individual events of the 'Man Bites Dog' type" [7, p. 179]. Furthermore, they highlight the psychological aspect that individuals who heavily rely on universal media[5] may develop "completely distorted perceptions of societal risks" [7, p. 179]. This type of impact on recipients, stemming from a scandalous perspective portrayed through schema-driven media contributions,[6] is extensively examined in scandal research.[7]

---

[3] For further reading on hacker ethics, refer to [5, pp. 27-38].
[4] Cf. the term displacement (ger. Verschiebung) as discussed in [6].

## 2.1. Mechanisms of Scandalization

According to Sebastian Pflügler and Philip Baugut, the term *scandal* in scandal research is widely defined as "a human, democratic, context-bound, and temporally limited phenomenon" [12, p. 309] wherein "the alleged norm violation of a known person, the scandalized, is revealed by a scandalizer and elicits outrage from the public" [12, p. 309]. The mechanisms employed in journalistic scandalization serve as a supplementary tool for examining the portrayal of hackers in media reporting, aiming to construct an understanding of media reality. Through an analysis of scandalization mechanisms, which are intrinsic to journalistic practices, it becomes possible to elucidate the process of frame-building, thereby facilitating an analysis of media frames related to the social figure of the hacker.[8] Grounded in media effects research, media reality can be apprehended as a secondary construct through the use of content analysis [8, p. 175].

Pflügler and Baugut delineate four types of scandal research in their systematic analysis, which they define as ideal types using an analytical framework [12, pp. 310-312]. The authors illustrate that four fundamental manifestations of scandal research can be typified: Firstly, societal moralism, wherein the outbreak is attributed to a norm violation [12, pp. 312-314]. Secondly, media-oriented constructivism, wherein the cause of the scandal's onset lies in the reaction of the media system, which is inherently interested in the construction itself [12, pp. 314-316]. Thirdly, strategic constructivism, wherein the impetus for the onset of a scandal is the reaction of the media system itself, implying the aspect of power concretely [12, pp. 316-318]. Fourthly, internet-oriented relativism, wherein the media channels and their societal impacts are precipitated by the audience or users themselves due to decontextualization in the Web 2.0 environment [12, pp. 318-320].

---

[5] Espey and Rudniger define mass media as television, radio, and the 'tabloid press' [7, p. 179]. For an overview of mass media as ubiquitous communication channels, refer to [9, p. 117]. For criticism regarding the origin of the term concerning internet research, see [10, p. 453].
[6] [11, pp. 27-49]; within academic literature, the term *frame* is used synonymously with *schema*: [8, p. 189].
[7] For the classification of types of scandal research, see: [12].
[8] For the concepts of frame-building and media frames, refer to [8, pp. 189-192].

The systematic typology of scandal research, with its specific distinguishing features, as outlined above, is comprehensively detailed in Hans Mathias Kepplinger's seminal work on the mechanisms of scandalization [11]. According to Kepplinger, scandals, from a social science perspective, exhibit six specific characteristics [11, pp. 27-28]. Broadly summarized, these are: Cause, Trigger, Self-interest, Awareness, Media presentation, Consequences. The cause (1) of scandals lies in "material and immaterial grievances" [11, p. 27]. Subsequently, these grievances are triggered (2) by an individual. Their actions are influenced by their own self-interest (3). Moreover, individuals are aware (4) of the consequences of their actions but choose not to act differently. Another characteristic is the media presentation (5), which portrays events "very intensively and largely uniformly" [11, p. 28]. As a result, there are consequences (6) for the social reality of those involved in the scandal, who are responsible for the grievances. The consequences, strongly demanded by the media public, are particularly evident in the demand for the perpetrators to be "held accountable" [11, p. 28]. This illustrates that scandals are fundamentally "not natural reactions to grievances but consequences of identifiable mechanisms of public communication" [11, p. 8].

In the subsequent analysis of the portrayal of the hacker's media reality, tools facilitating a linguistically analytical perspective become paramount. These tools aid in discerning the media frames that shape such a reality and coalesce in reporting. Among these tools, dramatization stands out as a prominent mechanism of scandalization. This attribute necessitates linguistic scrutiny, involving rhetorical devices, to fulfill the expected function [11, p. 64]. In this context, Kepplinger delineates seven types of dramatization in the process of scandalizing grievances: horror labels, associations with crime, defamation, suggestions of catastrophes, collage-like portrayals of catastrophes, serial scandalization, and visual exaggerations [11, pp. 66-67].

Horror labels are extreme terms used in reporting to describe damages or grievances, such as forest dieback, killer bacteria, catastrophe, or disaster [11, p. 66]. The type of criminal association characterizes norm violations as serious crimes or major breaches of general ethical principles, exemplified by terms like eavesdropping, constitutional breach, and killer [11, p. 66]. Defamation serves as the counterpart to horror labels in political and church scandals, and to criminal associations in industrial and environmental scandals, as seen in terms like disgusting baker and showy bishop [11, p. 66]. Catastrophe suggestion involves portraying conceivable maximum damages as imminent threats, though their actual unlikelihood is disregarded. Examples include terms like BSE, bird flu, swine flu, Chernobyl, and Fukushima [11, p. 66]. Catastrophe collage juxtaposes [11, p. 66] damages and dire conditions with extreme cases.[9] Serial scandalization portrays minor norm violations as part of a series of similar cases, creating the impression of a major grievance [11, p. 67], often triggered by the character of the scandalized individual.[10] Visual exaggerations involve the exaggerated depiction of norm violations, damages, or grievances through photos or films as particularly severe, dangerous, or frightening [11, p. 67].

## 2.2. Media Reality

In previous chapters, the term media reality was frequently used without providing a deeper, theory-oriented interpretation and explanation of the concept. Therefore, the following discussion will briefly examine the discourse and the meaning of the term within the German-speaking context to facilitate a better understanding. It's important to note that henceforth, the term *media reality* (ger. Medienwirklichkeit) will be used in accordance with its German meaning (see previous parenthesis). The term media reality linguistically comprises the root word *reality* and the attribute *media*. To initially differentiate, it is important to understand the following terms: In German the term reality (ger. Realität) represents a fundamental concept in the natural scientific perspective, rooted in the biological conception of perception. However, the term reality (ger. Wirklichkeit), semantically defined as actuality, is a constructed concept and serves to distinguish the term from pre-media reality. Therefore, media reality refers to the constructed media-based reality, encompassing the various levels of different individual media, each contributing to its unique representation of the world.

---

[9] Kepplinger refers to the edited volume by Gero Kalt and Michael Hanfeld: [13]. Additionally, he cites an example from a Spiegel TV program where AIDS viruses, mad cow disease, swine fever, and killer bacteria are assembled into a disaster collage: [11, p. 67].

[10] Prominent examples in Germany from the last ten years include the scandalization of individuals such as Peer Steinbrück, Franz-Peter Tebartz-van Elst, and Christian Wulff: [11, p. 67].

The following insights will provide reasons for this interpretation. It is important to note that the examination of the concept of reality and the associated analysis of the functioning of reality constructions are found in various disciplines, including philosophy, sociology, fine arts, language and literature studies, media studies, communication studies, political science, psychology, psychiatry, education, architecture, mathematics, physics, and biology [14, p. 4, 15, p. 129, 16, p. 11]. It becomes evident that the discourse on the concept of reality construction falls under the umbrella of constructivism. However, according to Paul Watzlawick, the term is unsuitable for several reasons, hence the recommendation to speak of reality research instead [17, p. 10]. Nevertheless, the term constructivism has become established in scholarly discourse [15, 16, 18, 19]. It should be noted in the context of the discourse on constructivism that there is no single constructivism; rather, there are various interpretations or versions, which nevertheless share a central commonality: the core problem of constructivism [14, p. 5]. This demonstrates that the issue is fundamentally interdisciplinary, focusing on observing or researching the processual understanding of the emergence of reality [14, p. 5], although different attempts at justification may be associated with it.

Despite the diverse arguments and justifications in their respective academic fields, the discourse on constructivism is held together and shaped by a series of interconnected postulates, leading figures, and conceptual frameworks [14, pp. 10-14]. Crucial to the entire discourse is "always the orientation toward the observer" [14, p. 11], and accordingly, the "center of attention [...] is no longer ontologically conceived what-questions, but epistemologically understood how-questions" [14, p. 11], which are at the "core interest of all constructivist authors" [14, p. 3]. It is precisely this aspect that constitutes the theoretical media-analytical foundations in media and communication studies. It is well known that the basis of all considerations regarding the relationship between media and reality is the question of what reality is, not whether such a reality exists [20, p. 33]. At this point, it becomes evident that media are not merely representational, mimetic, or reproductive, but rather generate their own world. As Knut Hickethier argues, the environment of humans represents the actual pre-media reality and is independent of reality [20, pp. 33-35]. In other words, the perception of humans and their environment constitutes the true reality. For media and communication sciences, this entails further differentiation because while referring to the biological construction of human perception and cognition is fundamentally accurate, it is of limited assistance in addressing the media. This is because it applies to all forms of reality construction and, as a result, serves as the foundation for the scientific examination of media reality, thus being perceived as universally valid in the discourse itself [14, 16].

In the field of media and communication studies, the adoption of constructivism has intensified an "already contentious foundational conflict between realists and relativists within the discipline" [19, p. 10]. Critique is particularly prominent in the context of journalistic practice and its representation, as well as inherent analysis: "There are concerns that the concept of objectivity may be diluted in epistemological discourse – and that such dilution could be construed as an invitation to forgery, manipulation, and deceit" [19, p. 12]. Consequently, media criticism would lose its foundation in any comparison between media reality and (absolute) reality [19, p. 12]. Additionally, a complicating factor arises from terminological ambiguity, identified by Bernhard Pörksen as the "problem of referential confusion" [19, p. 12]. This confusion often arises from conflating statements referring to absolute reality with those made within the confines of known limitations of knowledge, intended for the realm of everyday life and experiential reality [19, p. 12]. Following Watzlawick's framework, there exists a perceptual distortion between first-order reality and second-order reality.[11]

In light of concerns over the potential epistemological erosion of the foundation of journalistic practice, it is essential to contest the assertions made by critics of constructivism. Their apprehension that epistemological validation of arbitrariness would lead to a lack of established criteria for assessing the quality of media offerings overlooks significant insights from empirical communication research, journalism studies, political communication research, and media effects research. A culturalist epistemology, such as media and sociocultural constructivism, does not preclude the possibility of dichotomous evaluations of journalistic work. Rather, an examination of the research literature across various fields, including media studies, reception studies, media pedagogy, media effects research, journalism studies, research on political communication, and media sociology, reveals a fundamental challenge in source heuristic, significantly impacting the conception and construction of media reality [22, 23, pp. 221-262, 24, pp. 17-28, 25–28]. It is evident that this challenge extends beyond journalism into broader societal domains due to the ongoing digitalization and, notably, the process of media convergence, as evidenced by research on social media [29]. However, analytical approaches aimed at understanding the convergence effects of information technology systems on sociocultural dynamics remain fragmented across disciplines. Despite this, it is important to recognize that the epistemological focus varies markedly depending on the disciplinary context, contributing to the fragmented nature of current scholarly discourse.

In summary, it can be noted that the concept of media reality in media and communication science is understood as a dichotomous complementary term, as discussed in the previous examination of reality construction research [30, 31]. The constructivist nature of the term, characterized by the typification of realities, is influential, although various approaches exist to theoretically elucidate the relationship between media reality and reality (ger. Realität) [30, p. 225]. It is widely understood within the field that reality (ger. Wirklichkeit) is socially constructed, while also acknowledging the existence of intersubjectively understandable facts [32, pp. 101-102]. Consequently, drawing on the mechanisms of scandalization and dramatization, there appears to be a correlation between media reality and social reality, with their reciprocal influence contributing to the establishment of media reality as part of individual perception [11, pp. 195-197]. This has significant implications for the perception of pre-media reality through media reality, particularly evident in the portrayal of the hacker persona as discussed by Thomas Douglas in the context of Kevin Mitnick and Kevin Poulsen [1, pp. 175-219]. The question that arises now is: What is the journalistic portrayal of the hacker concept in Germany, and what public perception of the hacker persona can be observed?

---

[11] To understand the concepts of first and second-order reality, refer to: [21, pp. 142-143].

# 3. Media Representation: Data Analysis

Upon initial examination of the results from a longitudinal study encompassing the four highest-circulation German-language print and online media outlets during the period from January 1st to January 31st in the years 2017, 2018, 2019, and 2020,[12] it was observed that the term 'hacker' is predominantly associated with negative connotations in 80% of cases concerning IT security topics. Among these instances, 17% of articles were found to be unrelated to the specified topic, while only 3% exhibited positive connotations.[13] Notably, upon excluding non-specialized articles within the field of IT security, the prevalence of negative connotations increased significantly, encompassing 96% of all analyzed articles.

3.1. Explanation of Data Derivation for Longitudinal Analysis

To ensure transparency and comprehensibility of the dataset, circulation figures and sales numbers of the most widely circulated German-language online and print media were sourced from *the Informationsgemeinschaft zur Feststellung der Verbreitung von Werbeträgern* (IVW). Specifically, the Süddeutsche Zeitung, Frankfurter Allgemeine Zeitung, Die Welt, Handelsblatt, Die Zeit, Der Spiegel, Focus, and Bild Zeitung were identified among the top-ranking media outlets for the years 2017, 2018, 2019, and 2020 [33]. The weekly period from January 1st to January 31st was chosen for the analysis of relevant articles. This timeframe was selected due to its association with the beginning of a new year and to avoid the journalistic summer [news] hole (ger. Sommerloch), a period characterized by a scarcity of important political news during the summer vacation period [35], typically lasting from mid-June to mid-September in Germany. Additionally, the choice of this timeframe allows for easy accessibility of articles over an extended period. Articles hidden behind paywalls, denoted as paid content, were excluded to ensure unhindered access to the data. The online editions of the following print media outlets, including the Frankfurter Allgemeine Zeitung, Handelsblatt, and Die Welt, have continuous paywalls for entire monthly and yearly sections, which are fundamental for the longitudinal analysis. Consequently, these newspapers were excluded from the dataset to ensure transparency and comprehensibility. Otherwise, it would not be feasible to create a comprehensive cross-sectional survey with staggered data, rendering the longitudinal analysis incomplete. The remaining newspapers' online platforms allow for the chronological listing of articles containing the keyword 'hacker' through search queries. Unfortunately, this functionality is not available for the news magazine Focus, rendering it unsuitable for inclusion in the dataset.

For the longitudinal study, the Süddeutsche Zeitung and Bild Zeitung were chosen as national daily newspapers, Die Zeit as a national weekly newspaper, and Der Spiegel as a print news magazine. Twenty articles were selected for each month of January (beginning on January 1st) over four years, resulting in a total of 80 articles across the four chosen online media outlets for the selected monthly period. This corresponds to a total of 320 articles over four years, representing an ideal dataset in which all 320 articles are accessible. Furthermore, for each January month, the respective newspaper published precisely or more than 20 articles, and any potential paywall applies only to individual articles rather than the entire month. Considering these criteria, a refined dataset comprising a total of 301 articles was obtained.[14] The data

---

[12] To access sales figures and print circulation data, refer to the [33].
[13] See Appendix: [34].
[14] In the dataset (January 2018), there are a total of three empty positions for Bild Zeitung, as 19 articles were published from January 1st to January 31st, two of which were behind a paywall. In the dataset (January 2020),

attributes for each survey are categorized into nine characteristics: newspaper name, article title, publication date, semantic field, negative or positive connotation, possible keywords (related to connotation and semantic field), determination of whether the article is off-topic (regarding IT security), date of access, and the article's URL serving as the endpoint reference.[15]

## 3.2. Content Analysis

In the dataset, four components have been identified for conducting the longitudinal analysis, facilitating the determination of media frames based on linguistic criteria (see Figure 1):

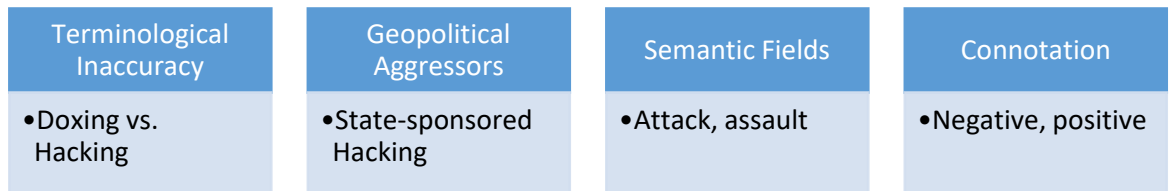

*Figure 1: The linguistic components of the content analysis.*

The primary issue is terminological inaccuracy, where incorrect terms are used despite the existence of socio-cultural terms that would accurately describe the situation. This initial content analysis confirms Espey and Rudinger's criticism, revealing a disparity between journalistic knowledge and ignorance. For instance, in early 2019, reports emerged of data belonging to hundreds of politicians being published. Initially labeled as a hacker attack, it was later revealed to be a case of doxing – the deliberate online publication of personal information about an individual with the intent to threaten, humiliate, or intimidate [For the term doxing, see, among others: 36, p. 199, 37, pp. 210-211, 38]. The personal data and information can typically be compiled by any individual through thorough research [36, 38]. Thus, doxing itself is not a form of hacking, although the means of obtaining information about a target bear resemblance to the implicit principles of OSINT (Open Source Intelligence)[16] methodology. However, the objective of doxing differs significantly due to the step of actively publishing the data. Secondly, the articles contain clear references to geopolitical aggressors, primarily categorized as state-sponsored hacking. Throughout the years covered in the dataset, Russia and Iran consistently emerge as prominent countries, with Saudi Arabia joining the geopolitical narrative as a presumed aggressor in 2020. This textual analysis reveals country-specific attributions in the articles, often seen in phrases combining adjectives like 'Russian' or 'Iranian' with terms such as 'hacking' or 'cyberattack.' Thirdly, it is important to consider the linguistic component of semantic fields, which reveals the semantic relationship of the term 'hacker' within the dataset. Here, listed in descending order of frequency within the dataset, are terms such as 'attack,' 'assault,' 'cyberwar,' 'fear,' and 'data.' This suggests that the term 'hacker' is predominantly associated with the semantic fields of 'attack,' 'assault' or 'cyberwar' in the coverage from 2017 to 2020, specifically in the month of January. At the fourth and final position, it is important to note the linguistic component of connotation. This pertains to a "general, social, and culturally established aspect of meaning" [40, p. 352] inherent in "a lexical expression, often carrying a pejorative undertone." [40, p. 352].

---

there are a total of 16 empty positions: Bild Zeitung published 15 articles, six of which were behind a paywall. Die Zeit published more than 20 articles, 12 of which were behind a paywall. See: Appendix: [34].
[15] Cf. Appendix: [34].
[16] According to the German Federal Intelligence Service (BND), this refers to "the systematic and targeted acquisition of publicly available information. It occurs across the entire spectrum of publicly accessible information channels," as quoted in: [39, p. 32].

For the content analysis of the dataset, a binary classification was chosen to attribute articles with either a negative or positive connotation. For example, when an article features the keyword 'hacker attack', it not only highlights the semantic field of attack but also indicates a clear negative connotation. Furthermore, in determining the connotation, we draw upon the characteristic of dramatization as one of the mechanisms of sensationalism. It is observed that out of the seven types of dramatizations (excluding visual exaggeration as it does not constitute a linguistic feature in itself), all six types are identified. From this, it can be inferred that the term 'hacker' in the selected articles is often placed in a semantic relation with negative word fields in multiple textual forms, thereby largely reinforcing the negative connotation. In detail, the analysis for the month of January 2017 reveals that the term 'hacker' carries a negative connotation a total of 64 times, accounting for 80% of the instances. Conversely, 5 articles portray a positive sentiment, comprising 6% of the corpus, while 11 articles are tangential to the subject matter, approximately representing 14% of the dataset. For a comprehensive presentation of the findings, please refer to Table 1 and Figure 2.[17]

|  | Jan | 2017 |  |  | Jan | 2018 |  |  | Jan | 2019 |  |  | Jan | 2020 |  |  |
|---|---|---|---|---|---|---|---|---|---|---|---|---|---|---|---|---|
| *Newspapers* | DZ | S | B | DS | DZ | S | B | DS | DZ | S | B | DS | DZ | S | B | DS |
| *Negative* | 16 | 13 | 18 | 17 | 13 | 13 | 17 | 14 | 19 | 19 | 20 | 19 | 10 | 10 | 6 | 17 |
| *Positive* | 2 | 0 | 2 | 1 | 0 | 1 | 0 | 2 | 1 | 0 | 0 | 0 | 0 | 0 | 1 | 0 |
| *None* | 2 | 7 | 0 | 2 | 7 | 6 | 0 | 4 | 0 | 1 | 0 | 1 | 5 | 10 | 2 | 3 |
| *Total Number of Articles* | 20 | 20 | 20 | 20 | 20 | 20 | 17 | 20 | 20 | 20 | 20 | 20 | 15 | 10 | 9 | 20 |

*Table 1: Detailed listing of the connotation of the term 'hacker' in the month of January (2017 to 2020). Abbreviations for newspapers are as follows: DZ: Die Zeit, S: Süddeutsche; B: Bild; DS: Der Spiegel.*

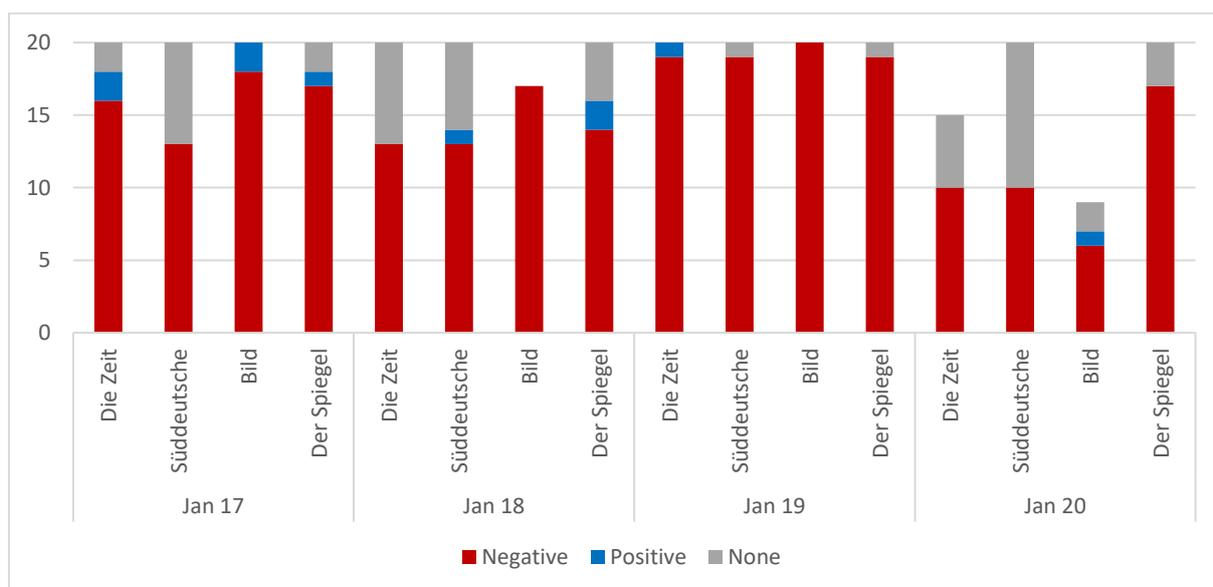

*Figure 2: Connotation of the term 'hacker' in the month of January from 2017 to 2020.*

---

[17] In subsequent data analysis, results are rounded to integers and, accordingly, presented without decimal notation.

In January 2018, the proportion of articles unrelated to the topic increased to 17, constituting 22% of the total, while three texts demonstrated positive connotations, representing a 4% share. However, articles with a negative connotation predominated, totaling 57, corresponding to 74% of the dataset. Transitioning to January 2019, a total of 77 articles were identified with a negative connotation, comprising 96% of all articles during this period. In contrast, three articles, distributed as one article positively utilizing the term 'hacker' (≈ 1%) and two articles unrelated to the topic (≈ 3%), were noted. Despite a significantly lower overall number of articles for January 2020 due to a refined dataset, the following distribution emerged from content analysis: 43 articles exhibited a negative connotation (≈ 67%), one article was positive (≈ 2%), and 20 articles were unrelated to the topic (≈ 31%) (for a graphical overview of the individual years, refer to Figure 3).

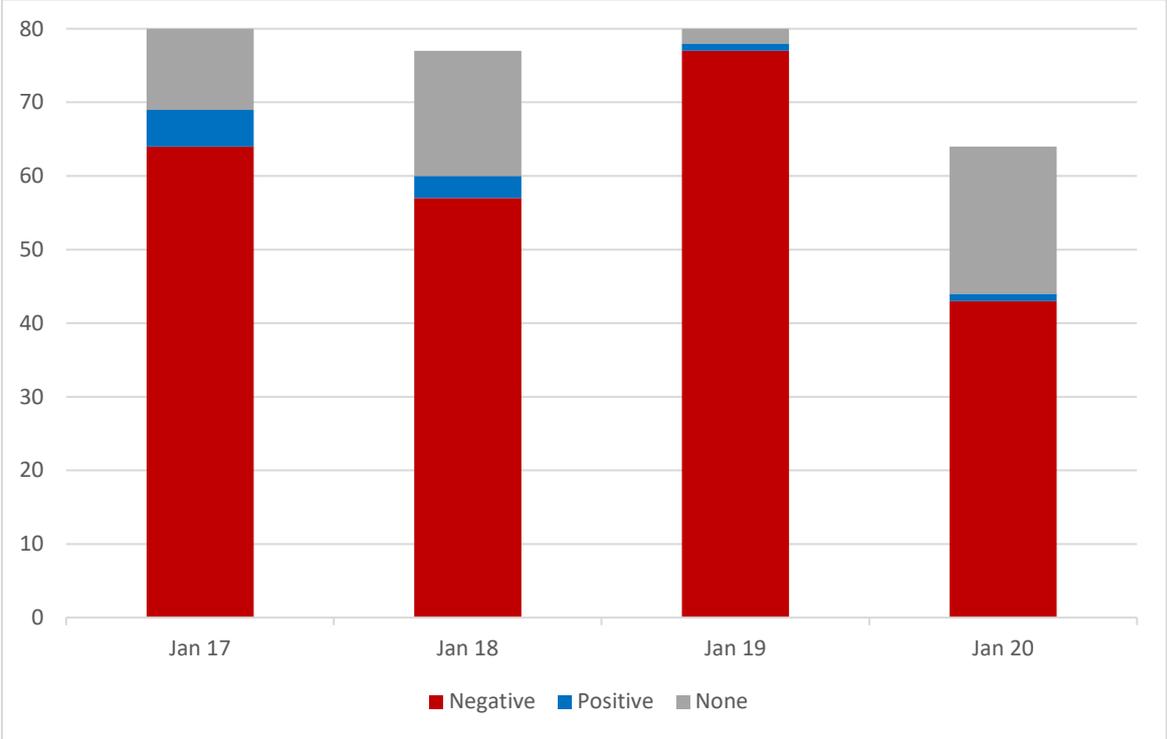

*Figure 3: Connotation of the term 'hacker' (total number of articles from 2017-2020).*

When projecting the data of the individually collected properties onto the overall depiction of the connotation of the term 'hacker' and consolidating the individual findings, the results of the longitudinal study of the four most widely circulated German-language print and online media, as outlined in Chapter 3.2, emerge. This signifies that, collectively, from 2017 to 2020, ten articles were published in January utilizing a positive attribution (≈ 3%). Additionally, 50 articles were unrelated to the topic (≈ 17%), while the term 'hacker' was negatively connoted 241 times, representing 80% of the total.

## 3.3. Media Frames as a Snapshot of Media Reality from 2017 to 2020

Based on the content analysis just conducted, it becomes evident that the media frame for the social figure of the hacker predominantly influences the perspective with a negative connotation. To further contextualize this media portrayal and ascertain potential implications, it is necessary to address questions regarding the effects suggested by the identified media frames and the media reality established based on the longitudinal study. At this juncture, the focus is not on recipients with preexisting thematic knowledge who are confronted with a corresponding media reality and may ultimately become, to some extent, part of the social reality of the individual. However, there is a need for a method to determine the effects and conceivable consequences without implicitly considering the multitude of variables that shape an individual's perception of reality. Operating under this premise, a brief thought experiment aims to facilitate the contextual interpretation of the results of the longitudinal study from a media effects perspective. For the thought experiment, we consider an ideally neutral archetype of a legally adult[18] individual, or person. By 'ideally neutral,' we mean that there exists no prior knowledge or preconceptions regarding the term 'hacker' or the subject of hacking. Furthermore, let us assume that the individual receives information solely based on the dataset of the aforementioned longitudinal analysis. This implies that the timeframe and information are confined to the months of January from 2017 to 2020 and to the journalistic media outlets of Süddeutsche Zeitung, Bild Zeitung, Die Zeit, and Der Spiegel. In this context, six aspects can be highlighted as effects that can be identified.

The first aspect to mention is the issue of validity, wherein the accuracy of the information is inaccessible to the recipient due to a lack of "the necessary information or expertise" [11, p. 40]. Applied to the realm of hacking, this becomes particularly evident in the reporting at the beginning of 2019 concerning the terms Doxing versus Hacking. Secondly, in reference to the thought experiment and the specified individual, an establishment effect can be observed, indicating that "a schema is formed in recipients, if at all" [41]. This is exemplified by the association of the term 'hacker' with the semantic fields of attack and assault, thereby establishing this media frame.[19] Building upon this, a perceived transformation effect arises, whereby an existing schema is "altered in the direction of the media frame" [41]. In the hypothetical scenario of an idealized individual, this phenomenon becomes apparent through the potential alteration of the neutral starting point. Here, the term 'hacker' may initially be perceived as neutral (value-free) by the individual lacking prior knowledge, but through media reporting, undergoes a connotative shift towards a negative or malevolent meaning.[20] Thirdly, it is important to highlight the ongoing reinforcement of the truth effect. The truth effect, defined as the tendency for individuals to attribute greater truthfulness to statements they have encountered previously compared to information they encounter for the first time [42], is particularly supported by the peculiarities of colleague orientation in journalism and the resulting degree of media self-referentiality [11, p. 80]. In the context of the media portrayal of hackers as observed in the longitudinal study: The more individuals are exposed to the portrayal of hackers as malevolent entities, the more likely they are to interpret this portrayal as accurate and truthful.[21] This continuous reinforcement occurs over time, commencing from January 2017 and continuing through the months of January from 2018 to 2020. The fourth aspect, problem simplification as a potential effect, again draws on reporting from 2019 (Topic: Doxing vs. Hacking). However, the consequence thereof is the establishment of media reality as part of one's own social reality, which applies to all the aforementioned aspects. The reason for this lies in the actual consequence for the individual, who may become disoriented due to problem simplification, as the extent of the threat cannot be accurately assessed [Cf. 11, pp. 195-197]. The fifth aspect to consider is the time factor of journalistic publications as a conceivable effect. Within the mechanisms of scandalization, this refers to the impact that "the longer a scandal lasts, the greater [...] the gap between what the majority believes to know and what can actually be known" [11, p. 197]. In the thought experiment, this would imply the emergence of an extended knowledge gap for the individual. Regarding the dataset of the longitudinal study, on one hand, this would pertain to factual knowledge, which, for example (see Doxing vs. Hacking), is not included in the reporting. On the other hand, it would encompass incorrect half-knowledge, which is part of the reporting. Lastly, the sixth aspect to mention is the increasing perpetuation effect. This phenomenon entails the continuous exposure of individuals to similarly connoted information. In the context of the thought experiment, this implies the solidification of the negative connotation. Therefore, the more negatively hackers and hacking are depicted, the more negatively this topic is perceived by the individual.[22] Consequently, it becomes evident that a media reality is established, consistently portraying a predominantly negative and adverse image of the subject matter.

---

[18] §2, §828 Abs. 3 (BGB), (ger. Bürgerliches Gesetzbuch), note: the BGB is the main body of civil law in Germany and regulates private law matters such as contracts, property, family law, and inheritance.
[19] Examples of the effects on the sociocultural level, see: [11, pp. 40-41].
[20] This type of media effect can also be observed in the mechanisms of scandalization, see: [11, p. 47].
[21] On the potential effects in journalism, cf. [11, p. 96].

## 3.4. New Potential Risks and Societal Challenges

The pseudo-threat associated with the social figure of the hacker, often distorted in journalism to instill fear of new technology, seems to have increasingly shifted in recent years from depicting individuals to portraying state aggressors. However, while significant new potential dangers cannot be distinctly identified solely based on the longitudinal study, three points emerge when considering the aforementioned geopolitical perspective. These include states themselves,[23] their intelligence agencies, and their respective militaries.[24] These three actors are increasingly emphasized in the media regarding cyber warfare,[25] reflecting the growing convergence of digital and conventional forms of warfare [45, p. 10]. Constance Kurz and Frank Rieger clarify this aspect by examining the current situation and identifying four indicators of warfare activities: "the significant financial investments in digital attacks, the professionalization of weapon construction, the employed strategies and tactics, and the involvement of military entities as active participants" [45, p. 10]. Active dissemination of disinformation also plays a pivotal role as a component of subversive warfare [43, 87–105, here esp. 91–92, Cf. 45, pp. 201-227]. In this context, journalism faces complex dilemmas [45, pp. 226-227], particularly concerning the challenge of discerning motives behind data leaks and the associated information presented to journalists [45, p. 227]. For society, it is anticipated that two fundamental issues will persist regarding the media portrayal. Firstly, concerning the depiction of hackers, unless there is a terminological shift at the journalistic level, hackers will continue to be portrayed as an omnipresent negative universal term, symbolizing danger and threat in the journalistic portrayal of new technology. Secondly, a problem becomes evident concerning the acquisition of knowledge based on information from relevant media sources. This implies that individuals lacking knowledge in IT security or IT in general are likely to encounter a concerning dichotomy when faced with processed information: media-derived half-knowledge juxtaposed with ignorance, understood here as a state of not knowing.

---

[22] A similar effect is also described by Kepplinger regarding politics and politicians in media representation: [11, p. 217].
[23] In the relevant literature, attacks by and through states are referred to as 'nation-state attacks', cf. [43].
[24] Detailed examples concerning the countries China, Russia, Iran, the USA, and North Korea can be found in: [43, pp. 4-34].
[25] For a critique of the term cyberwar, see: [2, pp. 437-438, 44, p. 68].

## 3.5. Youth and IT Security

In the context of educational science, the portrayal of hackers in the media and the associated theme of hacking prompt the question: How do adolescents address the issue of IT security? According to the Digital Barometer 2019, published by the Federal Office for Information Security (BSI) and the Police Crime Prevention of the Federal States and the Federal Government (ProPK), younger individuals, including adolescents, in Germany tend to exhibit more careless or reckless online behavior, resulting in a higher incidence of falling victim to cybercrime [46]. The authors attribute this, among other factors, to the fact that 26% of respondents aged 16 to 29 do not promptly implement recommendations for internet security.[26] Turning the perspective from Germany to the international sector, there appears to be a level of risk awareness among adolescents, according to Arkaitz Lareki, Juan Ignacio Martínez de Morentin, and their colleagues [49]. The findings of their study suggest that adolescents (aged 9 to 16 years) perceive risks related to computer security and, particularly, the handling of private data concerning data and photo publication more strongly than previously assumed. However, it's worth noting that the data for the study were collected through a survey of 1486 adolescents in the Basque Country and Navarre regions in northern Spain [49, p. 397].

In a representative survey of adolescent computer users in Germany, Flavius Kehr, Tobias Rothmund, and their colleagues examined the relationship between adolescents' protective behavior and five influencing factors to identify relevant predictors of such behaviors [50]. These factors are categorized as follows: Firstly, the factor of generalized trust [50, p. 304], defined as "the generalized readiness to trust a person" [51]. Secondly, the assessment of uncertainty regarding risk perception [50, p. 304]. Thirdly, self-assessment of expertise in computer handling [50, p. 304]. Fourthly, knowledge of basic IT security measures [50, pp. 304-305], and fifthly, parental control over usage [50, p. 305]. Subsequently, the authors present three fundamental findings based on statistical analysis. On one hand, it is evident that "there is relatively little knowledge regarding security-related measures" [50, p. 305]. On the other hand, a "linear trend was identified, indicating that knowledge increased linearly with grade level and type of school" [50, p. 305]. Furthermore, a strong heterogeneity was observed regarding the correlation between computer ownership and knowledge of potential IT security measures [50, p. 305]. Consequently, Kehr, Rothmund, and their colleagues conclude that "the degree to which a person exhibits protective behavior significantly increases with greater computer expertise, risk assessment, knowledge, parental control, and generalized trust" [50, p. 306].

In conclusion, it is evident that the studies cited herein consistently underscore the importance of IT knowledge, both broadly and specifically in relation to IT security, within the realm of education. These discussions frequently highlight deficiencies in IT literacy and knowledge, the lack of specialized IT skills, and the resulting dichotomy between those possessing knowledge and those without. As such, there is an implicit call throughout for the establishment of a foundational understanding of IT security as a fundamental educational requirement.

---

[26] [46] Note: The study does not transparently outline causal analysis methods, making it necessary to point out that no references to structure-testing procedures are provided. Consequently, the claim of causality in relation to a ratio or interval scale appears questionable, as the statistical evidence presented in the publication is insufficient to substantiate it. Nevertheless, a correlation between age and the implementation of IT security recommendations seems plausible. For foundational concepts in multivariate analysis methods, see [47, pp. 4-15]; for the issue and the relationship between causality and correlation in descriptive statistics, see [48, pp. 90-91].